\documentclass{emulateapj}
\usepackage{epsfig,graphics}
\usepackage[colorlinks=true, citecolor=blue, pagecolor=blue,
  linkcolor=blue, menucolor=blue, urlcolor=blue,
  linkbordercolor={0 0 1}, frenchlinks=True, breaklinks]{hyperref}

\begin{document}
\newcommand{\Ha}{H$\alpha$}
\newcommand{\Hb}{H$\beta$}
\newcommand{\OIII}{[O \textsc{iii}]}
\newcommand{\OII}{[O \textsc{ii}]}
\newcommand{\NII}{[\ion{N}{2}]}
\newcommand{\zphotf}{z_{\rm phot}}
\newcommand{\zspecf}{z_{\rm spec}}
\newcommand{\zphot}{$\zphotf$}
\newcommand{\zspec}{$\zspecf$}
\newcommand{\Rcf}{R_{\rm C}}
\newcommand{\Rc}{$\Rcf$}
\newcommand{\zband}{$z$\arcmin}

\newcommand{\galex}{{\it GALEX}}
\newcommand{\spitzer}{{\it Spitzer}}
\newcommand{\nuv}{{\it NUV}}
\newcommand{\fuv}{{\it FUV}}
\newcommand{\EBV}{$E$($B$--$V$)}

\newcommand{\mm}{$\mu$m}
\newcommand{\Msun}{M$_{\sun}$}

\newcommand{\iyr}{yr$^{-1}$}
\newcommand{\vMpc}{Mpc$^{-3}$}
\newcommand{\ergs}{erg s$^{-1}$}
\newcommand{\sbzk}{sBzK}

\newcommand{\rNB}{{\rm NB}}

\newcommand{\za}{0.40}
\newcommand{\zb}{0.49}

\newcommand{\AHa}{A({\rm H}\alpha)}

\newcommand{\NBa}{2512}  
\newcommand{\NBb}{1666}  
\newcommand{\speca}{326} 
\newcommand{\specb}{102} 
\newcommand{\specga}{232} 
\newcommand{\specgb}{73}  
\newcommand{\specua}{23} 
\newcommand{\specub}{6}  

\newcommand{\specHaa}{81}   
\newcommand{\specOIIIa}{78} 
\newcommand{\specOIIa}{42}  
\newcommand{\specLyaa}{21}  
\newcommand{\specHab}{47}   
\newcommand{\specOIIIb}{12} 
\newcommand{\specOIIb}{2}   
\newcommand{\specLyab}{1}   

\newcommand{\Ndual}{241}     
\newcommand{\Ndualz}{80}     
\newcommand{\Ndualbad}{nine} 
 
\newcommand{\Na}{401}    
\newcommand{\Nb}{249}    

\newcommand{\Nphotza}{266} 
\newcommand{\Nphotzb}{127} 

\newcommand{\Nspecbalmera}{73}  
\newcommand{\Nspecbalmeraa}{30} 
\newcommand{\Nspecbalmerab}{16} 
\newcommand{\Nspecbalmerac}{9}  

\newcommand{\Nspecbalmerb}{47}  
\newcommand{\Nspecbalmerba}{26} 
\newcommand{\Nspecbalmerbb}{10} 
\newcommand{\Nspecbalmerbc}{4}  

\defcitealias{ly07}{L07}
\defcitealias{hopkins}{H01}
\defcitealias{kennicutt09}{K09}

\submitted{Received 2012 January 13; accepted 2012 February 1}
\title{Dust Attenuation and H$\alpha$ Star Formation Rates of $z\sim0.5$ Galaxies}

\author{Chun Ly,\altaffilmark{1,9}
  Matthew A. Malkan,\altaffilmark{2},
  Nobunari Kashikawa,\altaffilmark{3,4}
  Kazuaki Ota,\altaffilmark{5}
  Kazuhiro Shimasaku,\altaffilmark{6,7}
  Masanori Iye,\altaffilmark{3} and
  Thayne Currie\altaffilmark{8}}
\shorttitle{Dust Reddening and \Ha\ SFRs}
\shortauthors{Ly et al.}
\altaffiltext{1}{Space Telescope Science Institute, Baltimore, MD, USA;
  chunly@stsci.edu}
\altaffiltext{2}{Department of Physics and Astronomy, UCLA, Los Angeles, CA, USA}
\altaffiltext{3}{Optical and Infrared Astronomy Division, National Astronomical
  Observatory, Mitaka, Tokyo, Japan}
\altaffiltext{4}{Department of Astronomy, School of Science, Graduate University for
  Advanced Studies, Mitaka, Tokyo, Japan}
\altaffiltext{5}{Department of Astronomy, Kyoto University, Kyoto, Japan}
\altaffiltext{6}{Department of Astronomy, School of Science, University of Tokyo,
  Bunkyo, Tokyo, Japan}
\altaffiltext{7}{Research Center for the Early Universe, School of Science,
  University of Tokyo, Tokyo, Japan}
\altaffiltext{8}{NASA-Goddard Space Flight Center, Greenbelt, MD, USA}
\altaffiltext{9}{Giacconi Fellow.}

\begin{abstract}
  Using deep narrow-band and broad-band imaging, we identify 401 $z\approx0.40$
  and 249 $z\approx0.49$ H$\alpha$ line-emitting galaxies in the Subaru Deep Field.
  Compared to other H$\alpha$ surveys at similar redshifts, our samples are unique
  since they probe lower H$\alpha$ luminosities, are augmented with
  multi-wavelength (rest-frame 1000\AA--1.5$\mu$m) coverage, and a large fraction
  (20\%) of our samples has already been spectroscopically confirmed. Our
  spectra allow us to measure the Balmer decrement for nearly 60 galaxies
  with H$\beta$ detected above 5$\sigma$. The Balmer decrements indicate an
  average extinction of $A({\rm H}\alpha)=0.7^{+1.4}_{-0.7}$ mag. We find that the
  Balmer decrement systematically increases with higher H$\alpha$ luminosities and
  with larger stellar masses, in agreement with previous studies with sparser
  samples. We find that the SFRs estimated from modeling the spectral energy
  distribution (SED) is reliable---we derived an ``intrinsic'' H$\alpha$
  luminosity which is then reddened assuming the color excess from SED modeling.
  The SED-predicted H$\alpha$ luminosity agrees with H$\alpha$ narrow-band
  measurements over 3 dex (rms of 0.25 dex).
  We then use the SED SFRs to test different statistically-based dust corrections
  for H$\alpha$ and find that adopting one magnitude of extinction is inappropriate:
  galaxies with lower luminosities are less reddened. We find that the
  luminosity-dependent dust correction of Hopkins et al. yields consistent results
  over 3 dex (rms of 0.3 dex). Our comparisons are only possible by assuming that
  stellar reddening is roughly half of nebular reddening. The strong correspondence
  argue that with SED modeling, we can derive reliable intrinsic SFRs even in
  the absence of H$\alpha$ measurements at $z\sim0.5$.
\end{abstract}



\section{INTRODUCTION}\label{1}
Dust extinction is the largest source of uncertainty in currently used
measurements of the properties of galaxies at substantial redshifts.
Reddening strongly effects the blue/UV continuum of galaxies upon
which most star formation rates (SFRs) are estimated
\citep{colbert06,ly09,goto10}.
Its effects are important across a wide range of redshifts
spanning much of cosmic time \citep{hicks02,shim09,newha}.
Although several methods have been used to correct for dust extinction, it
is still unclear how reliable each of them is, and under what circumstances.
Some are rough statistical estimates to be applied to large samples, such
as the common simple assumption of a constant one magnitude of extinction
for \Ha. This is an average, derived from studying the Balmer decrement of
local galaxies \citep{kennicutt92}. But there is significant scatter, so
that any single value will not apply to many individual galaxies.
Local studies have used UV, \Ha, and/or infrared measurements to
derive wavelength-dependent extinction or estimate dust-corrected SFRs
\citep{calzetti94,meurer99,buat02,salim07}.
Also, studies have found that the dust reddening properties are correlated
with galaxy's properties \citep[e.g., luminosity,][]{wang96}.
As such, reddening corrections should be measured for individual galaxies,
from spectroscopy of their gas ionized by young stars. 

The ``gold standard"  estimator is the Balmer decrement--\Ha/\Hb, which gives
a reddening assuming a simple thin dust screen covers the \ion{H}{2}
regions in a given galaxy.
This measured gas extinction in a galaxy might then be applied to its
starlight, assuming that it is the same, or alternately half as much
\citep{calzetti00}.
However, Balmer decrement measurements require rest-frame optical spectra,
which are especially difficult to obtain in high-redshift galaxy surveys.

In this paper, we present new data on the SFRs and dust reddening for $\sim$400
\Ha-selected galaxies at $z\sim0.4$--0.5. The sample is obtained from
the Subaru Deep Field \citep[SDF;][]{kashik04} with deep broad-band and
narrow-band (NB) imaging from Subaru's prime focus imager, Suprime-Cam
\citep{miyazaki02}.
Throughout this paper, (1) all distance dependent measurements adopt a flat
cosmology with $\Omega_{\Lambda}=0.7$, $\Omega_M=0.3$, and
$H_0=70$ km s$^{-1}$ Mpc$^{-1}$,
(2) a \cite{chabrier03} initial mass function is assumed, and (3) we
adopt a \cite{calzetti00} dust reddening formalism.


\section{The Sample}\label{2}
\begin{figure}
  \epsscale{1.1}
  \plotone{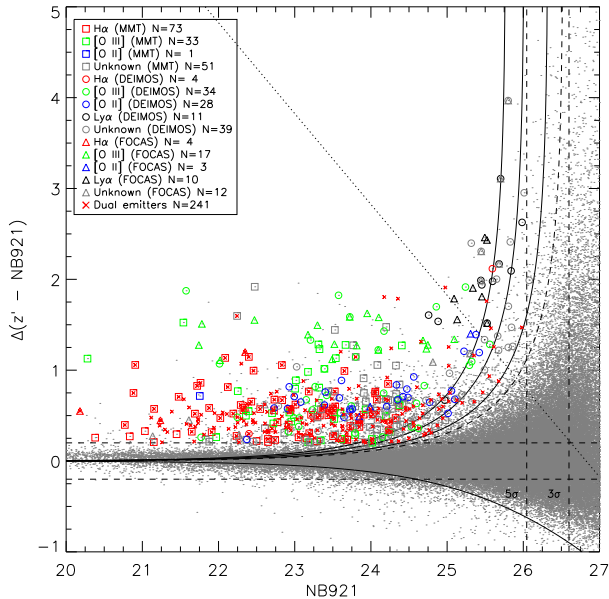}
  \caption{Continuum-corrected $z$\arcmin--NB921 colors for NB921 emitters.
    Solid (dashed) curves show detections that are significant at the 5.0,
    4.0, and 3.0$\sigma$ (2.5$\sigma$). \NBa\ galaxies meet the minimum
    0.2 mag excess (horizontal dashed line) and is above a 3$\sigma$
    detection in $z$\arcmin--NB921. Spectroscopically targeted sources
    are overlaid with symbols representing different instruments: DEIMOS
    (circles), Hectospec (squares), and FOCAS (triangles). \Ha, \OIII,
    \OII, Ly$\alpha$, and unidentified sources are distinguished by red,
    green, blue, black, and grey colors, respectively. Sources above the
    diagonal line are undetected in the $z$\arcmin-band at 3$\sigma$.
    Vertical lines refer to 3$\sigma$ and 5$\sigma$ NB921 limits.}
  \label{z_NB}
\end{figure}

Our new analysis is based on two unique samples of emission-line galaxies
selected in SDF by their excess flux in NB filters \citep[hereafter L07]{ly07}.
These are NB921 ($\lambda_{\rm C}=9196$\AA; ${\rm FWHM}=132$\AA) and NB973
($\lambda_{\rm c}=9755$\AA; ${\rm FWHM}=200$\AA).
Our selection of NB excess emitters uses the NB color--magnitude diagram
(illustrated in Figure~\ref{z_NB}), as in previous papers
(\citealt{fujita03}; \citetalias{ly07}, and references therein).
In total, \NBa\ NB921 and \NBb\ NB973 line emitters are identified, with
a minimum observed equivalent width (EW) of 32\AA\ (0.2 mag NB excess),
and 70\AA\ (0.25 mag NB excess), respectively.\\
%
Since the NB filters are not centered on the central wavelength of the
\zband-band, correction to the continuum fluxes, as derived by
extrapolating the $i$\arcmin--$z$\arcmin\ color, was applied to estimate
the flux at the wavelength of the NB filter. 
This color correction decreases the scatter in \zband--NB for non-NB excess
emitters (at bright magnitudes) by as much as $\sim$50\% for the NB973 sample,
and allows us to select genuine emission-line sources down to 0.25 mag excess. 
The correction also reduces the contamination from red galaxies without
emission lines, which are our greatest ``false positives.''
%
%
\subsection{Identification of \Ha\ Emitters}
Next we separate out the NB-excess sources whose NB emission is due to the
\Ha\ line, at redshifts of $z=0.401\pm0.01$ (NB921) and $z=0.486\pm0.015$
(NB973). In particular, these must be distinguished from other NB excesses
produced by other strong nebular emission lines, such as \OIII\ $\lambda$5007
and \OII\ $\lambda$3727 at higher redshifts.
Our preferred method of identifying the \Ha\ emitters is through
spectroscopic redshifts, where available (see colored symbols in
Figure~\ref{z_NB} and Section~\ref{spec_sec}).

\begin{figure*}
  \epsscale{0.5}
  \plotone{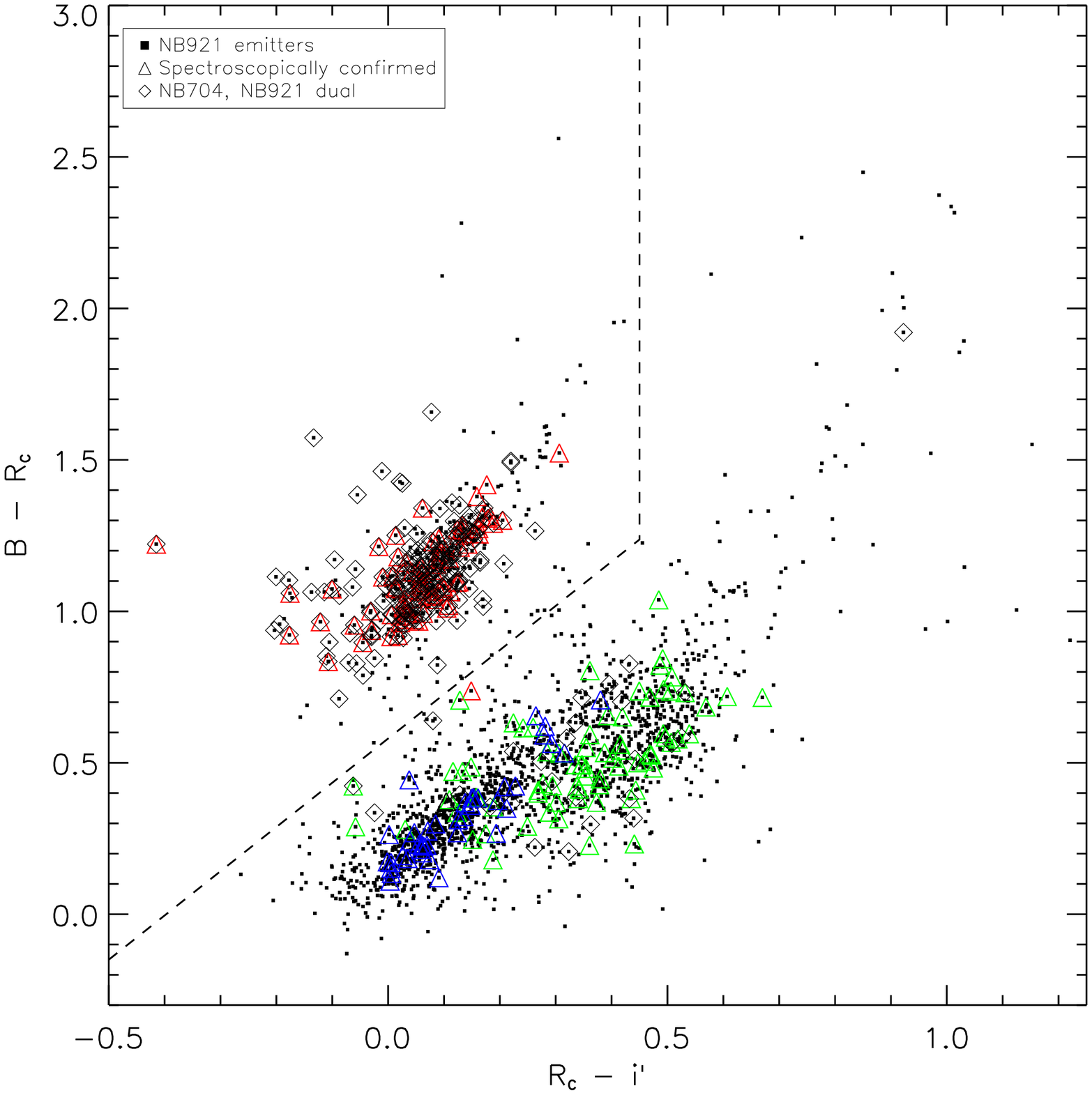}
  \plotone{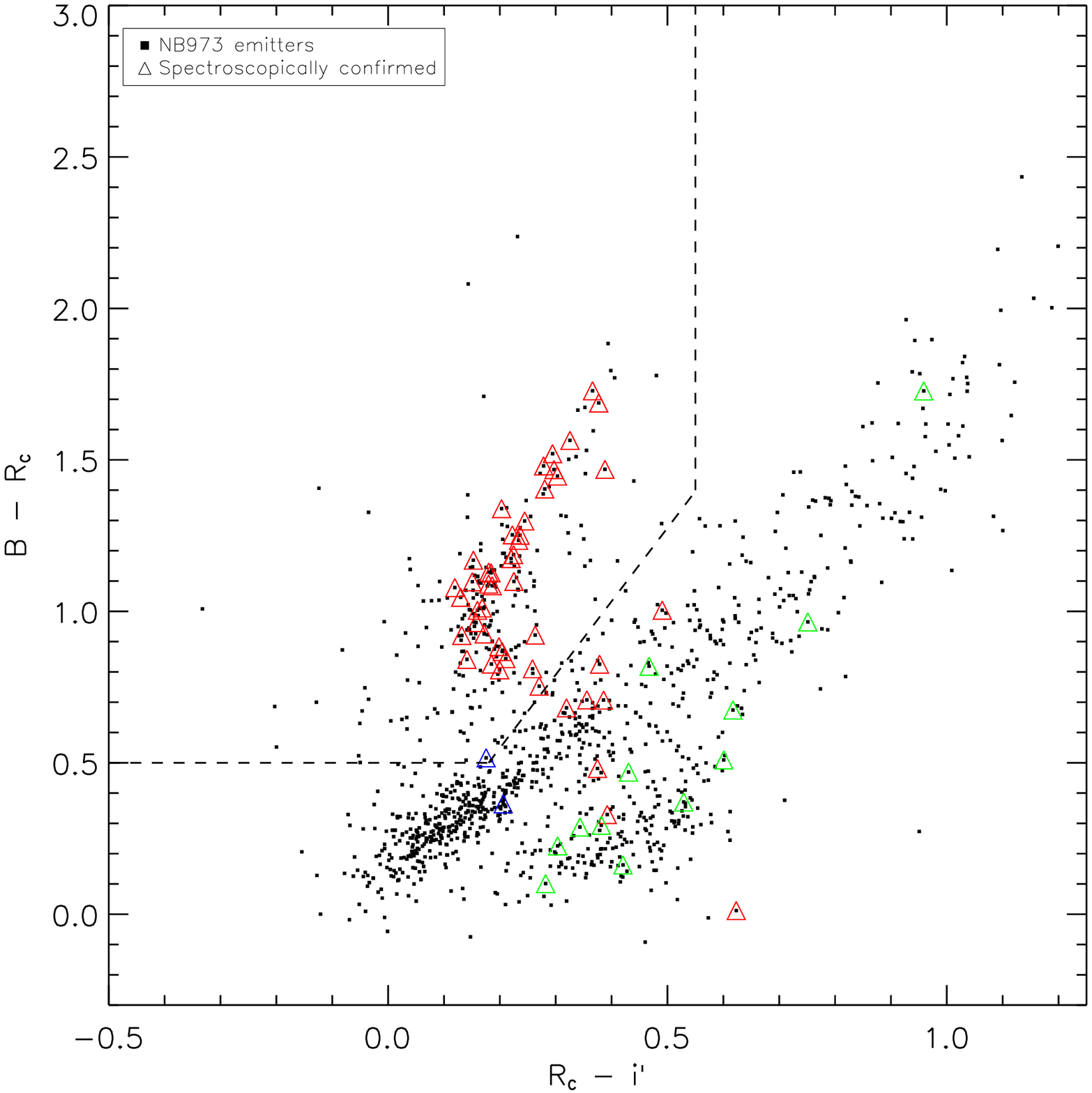}
  \caption{$B$--\Rc\ and \Rc--$i$\arcmin\ colors for NB921 (left) and NB973 (right)
    excess emitters. Sources detected above 5$\sigma$ in all three bands are only
    shown. Spectroscopically confirmed sources are shown as triangles, with their
    colors denoting their redshifts: \Ha\ (red), \OIII\ (green), or \OII\ (blue)
    in the NB filter. Dual NB704+NB921 emitters are shown as diamonds on the left
    panel. The majority of these dual emitters are at $z\approx0.4$ (\OIII\ and
    \Ha). Dashed lines show the color criteria to be classified as an \Ha\ emitter
    when spectroscopy is not available.}
  \label{BRRi}
\end{figure*}

For the remaining NB-emitting galaxies lacking spectroscopy, we use the
broad-band color technique developed in \citetalias{ly07} to identify
\Ha\ emitters.
We illustrate in Figure~\ref{BRRi} the $B$--\Rc\ and \Rc--$i$\arcmin\ colors
of our NB excess emitters.
There is a clear locus of sources with red $B$--\Rc\ and blue \Rc--$i$\arcmin\
colors. Color cuts (indicated by dashed lines in Figure~\ref{BRRi})
were chosen to separate these \Ha\ emitters from higher-redshift populations,
which lack a Balmer/4000\AA\ break in the $B$--\Rc\ color.
Such a selection was defined in our previous paper to yield a low
($\sim$3\%) interloper contamination rate \citepalias{ly07}.
The current larger sample strongly confirms the reliability of this
broad-band color classification, as we again find small interloper
fractions.
A total of \Na\ $z\approx\za$ and \Nb\ $z\approx\zb$ \Ha\ emitters
are identified. 
%
%
%
\subsection{Independent Confirmations of \Ha\ Reliability}
%
%
\subsubsection{Spectroscopy}
\label{spec_sec}
Using several optical spectrographs, we observed \speca\ NB921 and
\specb\ NB973 excess emitters.
Our NB973 spectroscopic observations are sparser, since that sample
became available much later. 
The majority (50\% and 78\% for the NB921 and NB973 samples, respectively)
of our spectra are from Hectospec \citep{fabricant05}, the multi-object
fiber-fed spectrograph for the MMT. We extend our spectroscopic sample
with Subaru's FOCAS \citep{kashik02}, and Keck's DEIMOS \citep{faber03}.
Our spectra typically have 1--2 hours of on-source integrations.
The acquisition and reduction of DEIMOS and FOCAS data have been
discussed in previous papers \citep{kashik06,ly07,kashik11}.
The MMT spectra were reduced following standard procedures with the
\textsc{hsred} reduction pipeline, and will be described further
in Ly et al. (2012, in preparation).

Among these spectroscopically targeted NB921-excess emitters, $\approx$71\%
(\specga) have reliable redshifts (based on at least one other emission line).
Similarly, 72\% (\specgb) of the NB973-excess emitters are spectroscopically
identified.
We note that these success rates are probably lower limits, since we do not
include \specua\ (an additional 7\%) NB921 and \specub\ (an additional 6\%)
NB973 sources that show single weak emission lines in their spectra which,
if they are real, are consistent with the redshifts predicted by an emission
line in the NB filter.
The numbers of spectroscopically confirmed \Ha, \OIII, \OII, and Ly$\alpha$ 
lines in NB921 (NB973) are \specHaa\ (\specHab), \specOIIIa\ (\specOIIIb),
\specOIIa\ (\specOIIb), and \specLyaa\ (\specLyab), respectively.
Thus, $\sim$20\% of our \Ha\ samples have been confirmed spectroscopically.

We note that Hectospec suffers from poor sensitivity and flat-fielding issues
redward of $\sim$8500\AA. While \Ha\ is not measured in these spectra, the
redshift determination is reliable due to detection of bluer emission lines,
generally \OIII, \Hb, and \OII.  And for 41 (20) cases with strong emission
lines, H$\gamma$ is also detected at $\gtrsim3\sigma$ ($\gtrsim5\sigma$)
significance.
Thus the spectroscopy provides several emission-line ratios, which are useful
for a number of astrophysical studies. In this Letter we concentrate on the
estimation of extinction from the Balmer decrement; we defer discussions of
gas metallicity measurements for these galaxies to a forthcoming paper
(Ly et al. 2012, in preparation).
%
%
\subsubsection{Dual Line Detection with NB Filter Pairs}
We have independent confirmations of the reality of a reliable
fraction of our NB emission lines, from imaging in a second NB filter.
Our NB704\footnote{$\lambda_{\rm C}=7046$\AA; ${\rm FWHM}=100$\AA.}
imaging of the SDF allows us to detect the \OIII\ emission at
$z\approx0.40$ while \Ha\ simultaneously enters the NB921 filter
\citepalias[this was first presented in][]{ly07}.
Other surveys \citep{hippelein03,sobral11,nakajima12} have exploited
similar dual line coincidences to confirm emission-line
galaxies at a particular redshift.

In SDF, a total of \Ndual\ ``dual emitters" were identified.
Among the dual emitters, \Ndualz\ were spectroscopically targeted,
and only \Ndualbad\ of them have unidentified redshifts. Among the
remaining 71 sources, 66 of them were \Ha\ emitters.
There are only two cases of dual \OII\ and \Hb\ at $z\approx0.89$.
Since these lines are weaker than \OIII\ and \Ha\ for
typical star-forming galaxies at these redshifts, 
it is not a surprise that the majority of spectroscopically
confirmed dual emitters are at $z\sim0.4$.
These dual emitters are shown as red crosses in Figure~\ref{z_NB}
and black diamonds in Figure~\ref{BRRi}; they further confirm that
the two-color selection for \Ha\ NB921 emitters is highly effective
(89\% meets the $B$\Rc$i$\arcmin\ color cuts).
Since the redshift ranges covered in \Ha\ by NB921 and in \OIII\
by NB704 only overlap by roughly 50\%, the large rate of dual emitters
indicates that nearly all \Ha\ emitters have \OIII\ emission, if only
it is searched for at the appropriate redshift.
%


\section{Observables for \Ha\ Emitters}\label{3}
%
%
\subsection{\Ha\ Luminosities}
The \Ha\ fluxes ($F_{{\rm H}\alpha}$) and luminosities ($L_{{\rm H}\alpha}$)
are derived (see \citetalias{ly07}) from the $z$\arcmin\ and NB filter
measurements, as follows:
\begin{eqnarray}
  F_{{\rm H}\alpha} & = & \Delta{\rm NB}\frac{f_{\rm NB}-f_{z}}{+1-(\Delta {\rm NB}/\Delta z)},~{\rm and}\\
  L_{{\rm H}\alpha} & = & F_{{\rm H}\alpha}\times(4\pi\,d_L^2),
\end{eqnarray}
where $f_X$ is the flux density in erg s$^{-1}$~cm$^{-2}$~\AA$^{-1}$ for band
``X'', $\Delta$'s are the FWHMs of the filters ($\Delta z$~=~955\AA), and
$d_L$ is the luminosity distance.
%
%
\subsection{Stellar Populations from SED Modeling}
We modeled the fifteen-band SED of each galaxy
to derive stellar masses, SFRs, dust reddening (\EBV), and stellar ages.
At $z\sim0.5$, the data span rest-frame 1000\AA--1.5\mm.
The spectral synthesis models are discussed further in \cite{ly11b}.
In brief, we use the FAST code \citep{kriek09}, with exponentially declining
$\tau$ star-formation history models from \cite{bc03}.
In these pure stellar-continuum broadband fits, the redshifts are accurately
fixed, either from the spectroscopic values when available, or from the presence
of \Ha\ in the NB bandpass.
If nebular emission lines were accounted for, we expect to obtain
somewhat younger less massive galaxy fits \citep{atek11}.


\section{Results}\label{4}
%
%
\subsection{Balmer Decrements}
A subset of our \Ha\ emitters with MMT spectroscopy\footnote{We only consider
  the great majority of our spectra, from the MMT, since their
  emission-line flux ratios are better determined.}
has significant detection of \Hb.
To obtain accurate emission-line fluxes from spectra, we calibrated
them using Hectospec observations of standards. 
To check the accuracy of our flux calibration, we compare the estimated
continuum fluxes in our spectra for 5000\AA--8000\AA\ to those from our
ancillary broad-band optical $BV\Rcf i\arcmin$ data. We find good
one-to-one agreement at the 20\% level. We correct for stellar \Hb\ absorption
by assuming a rest-frame EW of 2\AA.

To obtain the \Ha\ fluxes, two corrections must be applied. First, one must
correct the NB emission-line fluxes for contamination from \NII\
$\lambda$6548,6583. Local studies have found that the \Ha/\NII\ flux ratio
to be 2.3 \citep{kennicutt92,gallego97}.
However, in \citetalias{ly07}, we found from optical spectroscopy that many
of our \Ha\ emitters have very weak \NII\ emission. This is expected, 
as our galaxy population probes lower masses ($\approx10^{8.3\pm0.8}$ \Msun),
and the mass--metallicity relation would suggest weaker \NII\ emission.
In our same MMT spectroscopic observations, we obtained spectra of 
19 \Ha\ emitters at $z\sim0.25$ from another NB filter.
Aside from redshift, these galaxies are no different from those at $z\sim0.5$
in terms of luminosity, stellar mass, and \Ha\ fluxes.
For these galaxies we measured an average \Ha/\NII(6548+6583) ratio
of 5.6$\pm$2.0. We assume that this average applies to all our NB measurements. 

Second, the NB filters do not have perfectly rectangular transmissions, and as
a result, the measured emission-line fluxes can be underestimated when an
emission line falls away from the center of the NB filter. To correct for
this effect, we generated model spectra consisting of a flat continuum with
an added emission line of various strengths at the location that is dictated by the
spectroscopic redshift. 
We convolved these spectra with our filters' throughputs to measure the expected
observed $z$\arcmin--NB colors. When compared to the measured $z$\arcmin--NB colors
from our imaging, we are able to determine the intrinsic EW of the emission line,
and hence the intrinsic emission-line flux when scaled by the continuum flux.

The combination of spectroscopy and NB photometry
allows us to estimate Balmer decrement for individual galaxies.
Among \Nspecbalmera\ NB921 (\Nspecbalmerb\ NB973) \Ha\ emitters with MMT
spectroscopy, we have \Nspecbalmeraa\ (\Nspecbalmerba) 5$\sigma$ \Hb\ detections,
\Nspecbalmerab\ (\Nspecbalmerbb) 3--5$\sigma$ \Hb\ detections, and \Nspecbalmerac\
(\Nspecbalmerbc) 2--3$\sigma$ ``detections" which we show 3$\sigma$
lower limits on the Balmer decrements.

\begin{figure*}[htc]
  \epsscale{0.5}
  \plotone{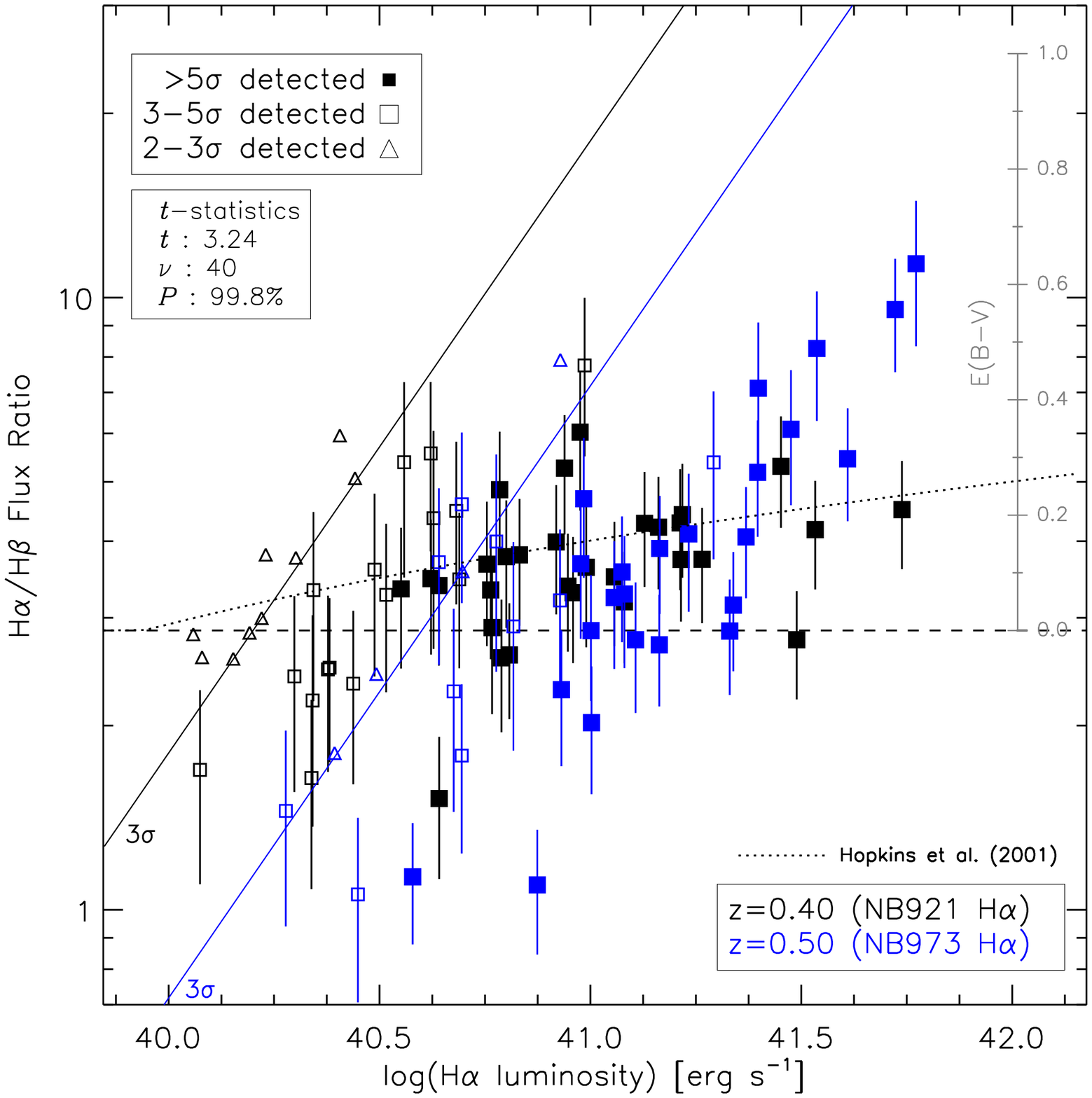}
  \plotone{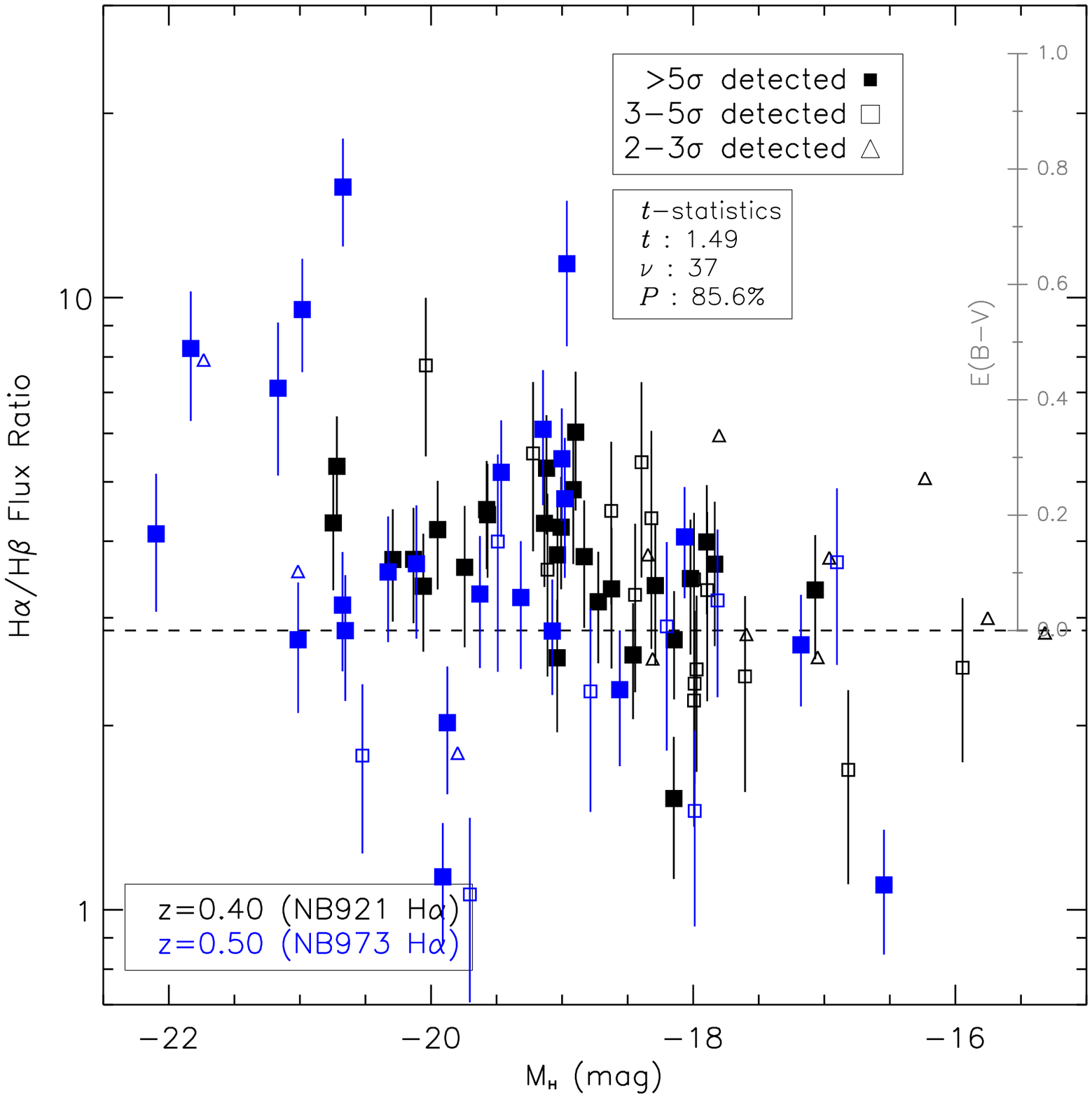}
  \caption{Balmer decrement as a function of observed \Ha\ luminosity
    (left) and $H$-band absolute magnitude (right).
    Black and blue symbols represent $z\sim0.4$ (NB921) and $z\sim0.5$
    (NB973) \Ha\ emitters. Filled and open squares refer to sources
    with $>5\sigma$ and 3--5$\sigma$ detections in the \Hb\ fluxes while
    triangles represent 3$\sigma$ upper limits on the \Hb\ fluxes for
    sources ``detected" at 2--3$\sigma$. The \citetalias{hopkins} \Ha\
    luminosity-dependent dust reddening relation is overlaid on the
    left panel (dotted line). \Hb\ 3$\sigma$ upper limit fluxes
    are shown by the solid black and blue lines. The stellar
    \EBV\ scale is shown.}
  \label{spec_compare}
\end{figure*}
In Figure~\ref{spec_compare}, we compare the Balmer decrements for these
galaxies against their \Ha\ observed luminosities and absolute $H$-band
magnitudes, $M_H$.
We find a correlation between the Balmer decrement and \Ha\ luminosity
(99.8\% significance), which is qualitatively consistent with those
derived for $z\sim0$ galaxies by \cite{hopkins}
(hereafter \citetalias{hopkins}; see Equation~\ref{ext}).
We determine that dust reddenings are systematically larger
in the more massive galaxies, but the correlation is
weak (85.6\% significance). Other studies \citep[e.g.,][]{garn10}
have also found a correlation between dust attenuation and stellar mass.

Although not illustrated, we did compare the \EBV\ inferred
from Balmer decrement against those estimated from SED
modeling. We find a weak correlation, and are limited
by significant Balmer decrement measurement
uncertainties and a small sample.
%
%
\subsection{Reliable Estimates on the Extinction-Corrected SFRs}
%
\begin{figure}[!htc]
  \epsscale{1.1}
  \plotone{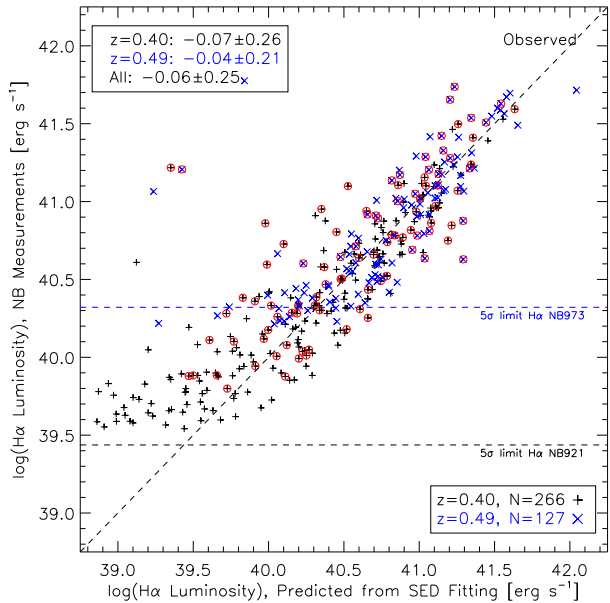}
  \caption{Comparisons between observed \Ha\ luminosities from NB measurements
    ($y$-axis) and estimates from SED modeling of the UV continuum ($x$-axis).
    The dashed line shows one-to-one correspondence. Spectroscopically
    confirmed emitters are indicated by red circles.}
  \label{Ha_pred}
\end{figure}

\begin{figure}[!htc]
  \epsscale{1.1}
  \plotone{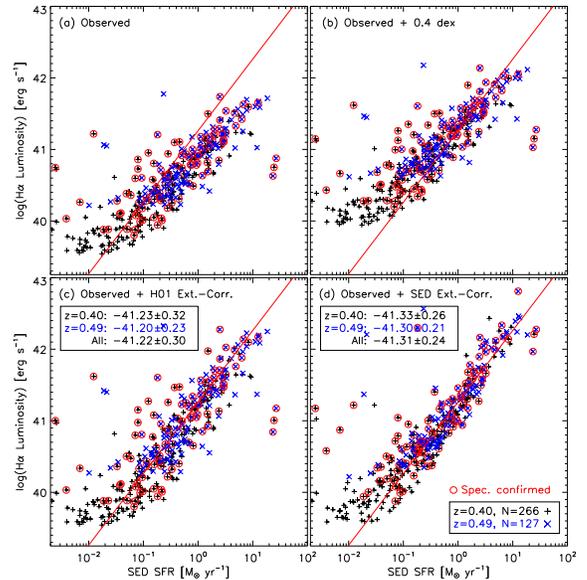}
  \caption{Comparisons between \Ha\ luminosities ($y$-axes) and SED-modeled
    SFRs ($x$-axes) for $z\sim\za$ (pluses) and $z\sim\zb$ (crosses)
    \Ha-selected galaxies. Each panel shows the \Ha\ luminosity
    with different assumptions about dust attenuation: (a) no correction,
    (b) $\AHa=1$ mag, (c) luminosity-dependent correction of \citetalias{hopkins},
    and (d) reddening from SED modeling. The \citetalias{kennicutt09} relation
    between \Ha\ luminosity and SFR is shown by the red lines with a
    logarithmic scaling value of -41.26.
    Spectroscopically confirmed emitters are indicated by red circles.}
  \label{SFR_plot}
\end{figure}

Since estimating \EBV\ from the Balmer decrement is
difficult, we pursue other methods to correct for dust attenuation.
We illustrate in Figure~\ref{Ha_pred} a comparison between the
{\it reddened} \Ha\ luminosity from NB observations and the
SED-{\it predicted} reddened \Ha\ luminosity:
\begin{eqnarray}
  L_{{\rm H}\alpha, {\rm SED}} &=& {\rm SFR}/K-0.4\AHa,{\rm~where}\\
  \label{AHa}
  \AHa &=& E(B-V)\times~k^{\prime}({\rm H}\alpha)\times1.85,{\rm~and}\\
  \nonumber
  k^{\prime}({\rm H}\alpha)&=&3.33.
\end{eqnarray}
The factor of 1.85 is for the ratio between stellar and nebular
reddening: \EBV$_{\rm gas}$ = 1.85\,\EBV\ \citep{calzetti00},
and $K$ is the conversion constant for the \citet[hereafter K09]{kennicutt09}
SFR--$L_{{\rm H}\alpha}$ relation:
\begin{equation}
  {\rm SFR(M_{\sun}~yr^{-1})}=5.49\times10^{-42} L_{{\rm H}\alpha}({\rm erg~s}^{-1}).
  \label{Krelation}
\end{equation}
Since the modeling was performed with only broad-band photometry,
the two axes are strictly independent of one another.
It illustrates that a nearly one-to-one correspondence
with an rms of 0.25 dex exists, and argues that the SFRs from SED
modeling (hereafter ``SED SFRs'') is reliable over 3 dex.
Because the model fits to SFR and reddening are primarily influenced
by the strength and slope of the UV continuum, the SED SFRs are
essentially dust-corrected UV SFRs.
This comparison also supports the idea that dust extinction
is lower for starlight than for gas, 
consistent with \cite{calzetti00}. Without a factor of 1.85,
Figure~\ref{Ha_pred} would show a statistical deviation of
0.10 dex (25\%) from one-to-one.

Using the SED SFRs as a reliable estimate of the intrinsic SFRs,
we illustrate in Figure~\ref{SFR_plot} the \Ha\ luminosity with
different statistical corrections for dust attenuation.
We limit our analyses to galaxies where their photometric redshifts
\citep[derived from EaZy;][]{brammer08} are consistent with the
NB redshifts. Excluded galaxies are generally fainter sources where
greater photometric uncertainties led to poorer fits.
The \Ha\ NB921 and NB973 samples are limited to \Nphotza\ and
\Nphotzb\ galaxies, respectively.

First, in Figure~\ref{SFR_plot}a, we show the {\it observed}
\Ha\ luminosity on the $y$-axis. The majority of sources clearly
fall below the \citetalias{kennicutt09} relation (red lines),
as expected since these luminosities have not yet been de-reddened.
Next we adopt three possible reddening corrections.
The first is a uniform $\AHa=1$ mag of extinction. This is shown in
Figure~\ref{SFR_plot}b. Although the average correction is reasonable,
it is systematically too large for those with low luminosities.

In the second, shown in Figure~\ref{SFR_plot}c, we adopt the
\citetalias{hopkins} SFR-dependent reddening formalism recast in luminosity:
\begin{eqnarray}
  \label{ext}
  \log{(L_{\rm obs})} & = & \log{(L_{\rm int})}-2.360\\
  \nonumber
  &\times&\log{\left[\frac{0.797\log{(L_{\rm int})}-29.10}{2.86}\right]},  
\end{eqnarray}
where $L_{\rm obs}$ and $L_{\rm int}$ are the observed and intrinsic \Ha\
luminosities, respectively. Adopting this dust-reddening correction, we
find consistency with the SED SFRs over 3 dex with an rms of 0.3 dex.

Finally in Figure~\ref{SFR_plot}d, we use individual estimates of
the $\AHa$ from modeling each galaxy SED (see Equation~\ref{AHa}).
Compared with the \citetalias{hopkins} correction, this
approach significantly reduces the scatter (rms of 0.24 dex). 
Figures~\ref{Ha_pred} and \ref{SFR_plot} illustrate the following results:
\begin{enumerate}
  \setlength{\itemsep}{1pt}
  \setlength{\parskip}{0pt}
  \setlength{\parsep}{0pt}
\item The assumption of $\AHa=1$ mag applies for
  observed \Ha\ luminosities of $\sim$10$^{40.5}$--10$^{42}$ \ergs\
  (observed SFRs above $\sim$0.2--10 \Msun\ \iyr), but is
  invalid for galaxies with SFRs $\lesssim$ 0.2 \Msun\ \iyr.
\item Dust reddening is greater in cases with higher \Ha\ luminosities,
  which have been suggested by previous studies. This is further shown
  in Figure~\ref{spec_compare}. 
\item The \Ha\ luminosity-dependent \citetalias{hopkins} relation succeeds
  in making the two independent estimates of SFR agree across
  three orders of magnitude ($10^{-2}$--10 \Msun\ \iyr) with
  an rms of 0.3 dex. This suggests that the behaviors of
  dust in star-forming galaxies have remained the same (relative to
  the \Ha\ luminosity) over the past five billion years.
  Similar findings by \cite{momcheva11} have been found, and
  extend the statement out to look-back time of $\sim$7 billion years
  ($z\sim0.8$).
\item With broad-band SED modeling (rest-frame 1000\AA--1.5\mm), we
  have determined reliable de-reddened SFRs that can predict the \Ha\
  luminosity over 3 dex, and suggests that SFRs derived from full
  SED modeling can be used as a reliable substitute when direct
  \Ha\ measurements are unavailable.
\item These comparisons also indicate that dust reddening for the
  gaseous component is more significant than the stellar component,
  and is in agreement with \cite{calzetti00}.
\end{enumerate}
%


\section{Conclusions}\label{5}
We used narrow-band imaging (at $\sim$9200\AA\ and $\sim$9750\AA) to
identify large samples of line-emitting galaxies in the Subaru Deep Field.
In our previous paper \citepalias{ly07} we developed broad-band color cuts
to isolate those produced by the \Ha\ line at $z\sim0.4$.  In this
paper we present new spectroscopy of $\gtrsim$400 NB excess emitters.
We confirm that the overwhelming majority of them are true line emitters
and that the previously used color selection technique is highly efficient
for identifying \Ha\ emitters.

We then find that the gas extinctions that we measure from the Balmer
decrement increase systematically with higher \Ha\ luminosity and with
higher stellar mass. In particular, the correlation with \Ha\ luminosity
is consistent with \citetalias{hopkins}.

Finally, we compare the de-reddened SFRs that adopt different
reddening corrections with the SFR obtained from modeling
the broad-band 1000\AA--1.5\mm\ SED.
We find that a constant reddening correction is not appropriate
for low luminosity galaxies, and that a correction which
increases with luminosity proposed by \citetalias{hopkins} yields
a good (0.3 dex) overall agreement with the modeled SFRs over 3 dex.
We also compare the observed \Ha\ luminosity from NB measurements
with those predicted when the SED SFRs are converted to an \Ha\
luminosity and reddened.
We find significant one-to-one agreement over 3 dex with an rms of
0.25 dex. Such finding suggests that broad-band far-ultraviolet
to near-infrared SED modeling can yield reliable SFRs for star-forming
galaxies at intermediate redshifts modulo direct \Ha\ measurements.

\acknowledgements
We thank Andrew Hopkins for helpful discussions, N. Caldwell with help
on designing Hectospec fiber configurations, R. Cool for additional
help using the MMT/Hectospec \textsc{hsred} reduction pipeline, and
the referee.
The authors wish to recognize and acknowledge the very significant
cultural role that the summit of Mauna Kea has always
had within the indigenous Hawaiian community.

{\it Facilities:} \facility{Subaru (Suprime-Cam)}, \facility{MMT (Hectospec)},
\facility{Keck (DEIMOS)}, \facility{Subaru (FOCAS)}, \facility{{\it GALEX}},
\facility{Mayall (MOSAIC, NEWFIRM)}, and \facility{UKIRT (WFCAM)}

\end{document}